\documentclass{aa}
\usepackage{graphicx}
\begin{document}
\title{Properties of galactic B[e] supergiants. IV. Hen3--298 and Hen3--303.}

\author{A.~S.~Miroshnichenko\inst{1,2,3}
\and K.~S.~Bjorkman\inst{1} \and M.~Grosso\inst{4} \and
K.~Hinkle\inst{5} \and H.~Levato\inst{4} \and F.~Marang\inst{6}}

\offprints{A.~S.~Miroshnichenko, \email{anatoly@physics.utoledo.edu}}

\institute{
 Ritter Observatory, Dept. of Physics and Astronomy, University of
     Toledo, Toledo, OH 43606-3390, USA
\and Max-Planck-Institut f\"ur Radioastronomie, Auf dem H\"ugel 69,
D-53121, Bonn, Germany
\and Central Astronomical Observatory of the
Russian Academy of Sciences at Pulkovo, 196140, Saint-Petersburg,
Russia
\and Complejo Astron\'omico El Leoncito (CASLEO), Casilla de
Correo 467, 5400 San Juan, Argentina \and National Optical
Astronomical Observatories, P.O. Box 26732, Tucson,
  AZ 85726--6732, USA
\and South African Astronomical Observatory, PO Box 9, Observatory
7935, South Africa
}

\date{received date, accepted date}

\titlerunning{Hen 298 and Hen 303}
\authorrunning{A.~S.~Miroshnichenko et al.}

\abstract{We present the results of optical and near-IR
spectroscopic and near-IR photometric observations of the
emission-line stars Hen3--298 and Hen3--303. Strong emission in the
H$\alpha$ line is found in both objects. The presence of Fe {\sc II}
and [O {\sc I}] emission lines in the spectrum of Hen3--298
indicates that it is a B[e] star. The double-peaked CO line
profiles, found in the infrared spectrum of Hen3--298, along with
the optical line profiles suggest that the star is surrounded by a
rotating circumstellar disc. Both objects also show infrared
excesses, similar to those of B[e] stars. The radial velocities of
the absorption and emission lines as well as a high reddening level
suggest that the objects are located in the Norma spiral arm at a
distance of 3--4.5 kpc. We estimated a luminosity of $\log
L/L_{\sun} \sim$5.1 and a spectral type of no earlier than B3 for
Hen\,3--298. Hen3--303 seems to be a less luminous B-type object
($\log L/L_{\sun} \sim$4.3), located in the same spiral arm.
\keywords{stars: emission-line -- B[e] stars: individual:
Hen\,--298, Hen\,3--303 -- techniques: spectroscopic, photometric --
circumstellar matter} }

\maketitle

\section{Introduction}\label{intro}

The B[e] stars are a heterogeneous group.  They are mostly of spectral
type B with forbidden emission lines in their optical spectra and large
IR excesses due to hot circumstellar (CS) dust (Allen \& Swings
\cite{as76}).  Although a number of B[e] stars have either been
identified as members of other known stellar groups (e.g., Herbig Ae/Be
stars and Proto-Planetary Nebulae) or suggested to have high
luminosities (B[e] supergiants, cf. Lamers et al. \cite{l98}), nearly
half of the originally selected 65 galactic objects remained, until
recently, unclassified.

Our studies of these unclassified objects resulted in identification
of a new distinct group of B[e] stars with IRAS fluxes suggestive of
a recent CS dust formation (hereafter B[e] stars with Warm Dust or
B[e]WD, Miroshnichenko et al. \cite{m02a}). Analysis of their
properties shows that they are neither pre-main-sequence nor
post-asymptotic giant branch (post-AGB) objects. Nearly a quarter of
B[e]WD were found to be detached binary systems, and a few more
objects were suspected in binarity. Furthermore, their wide range of
luminosities (2.5$\le \log$ L/L$_{\odot} \le$5.1) and the location
on the Hertzsprung-Russell diagram mostly within the main-sequence
suggest that dust formation near hot stars can be much more common
than originally thought. Most of the group members also exhibit very
strong emission-line spectra, which indicate either a strong ongoing
mass loss or the presence of a large amount of gas in their CS
environments.

B[e]WD can be distinguished from other B[e] stars by the IR spectral
energy distribution (SED). Analyzing the IRAS data of the originally
selected B[e] stars, Miroshnichenko and collaborators noticed that
10 objects had specific colours ($-0.5 \le \log (F_{25}/F_{12}) \le
0.1$, $-1.1 \le \log (F_{60}/F_{25}) \le -0.3$, where $F_{12}$,
$F_{25}$, and $F_{60}$ are the fluxes in the IRAS photometric bands
centered at 12, 25, and 60 $\mu$m, respectively) which correspond to
dust temperatures of $\ge$150--200 K.  Such colours are mainly
characteristic of late-type stars with CS dust (symbiotic binaries,
VV Cep binaries, Miras) and may indicate either the presence of a
cool companion or a compact dusty envelope without cold dust.  The
latter may be due to a recent dust formation process.
Cross-correlation of the IRAS Point Source Catalog (PSC) and the
catalog of galactic early-type emission-line stars (Wackerling
\cite{wack}) resulted in a finding of 11 more B-type stars with
similar IRAS colours.  Some of these objects were also selected by
Dong \& Hu (\cite{dong}), who searched for optical counterparts from
the same optical catalog to IRAS sources with strong IR excesses
($V-$[25] $\ge$ 8 mag, where $V$ is the visual magnitude from
Wackerling \cite{wack} and [25] is the magnitude, calculated from
the 25-$\mu$m IRAS flux).  Initial studies of B[e]WD, listed by Dong
\& Hu (\cite{dong}), were published by Miroshnichenko et al.
(\cite{m00}, \cite{m01}, \cite{m02b}).

The B[e]WD group includes a few high luminosity objects (e.g.,
\object{CI Cam}, \object{CPD$-52^{\circ}$9243}, \object{MWC 300},
\object{HDE 327083}), similar to B[e] supergiants of the Magellanic
Clouds (Zickgraf et al. \cite{z86}). The ``hybrid'' spectra of the
latter (P Cyg type spectral line profiles in the UV region in
combination with strong double- or single-peaked optical line
profiles) have been explained in the framework of a two-component
stellar wind model (dense and slow equatorial wind plus less dense
and fast polar wind). First attempts have been made to explain dust
formation around B[e] supergiants by Bjorkman (\cite{b98}) and Kraus
\& Lamers (\cite{kl03}).

Our results on galactic B[e] supergiant candidates show that their
luminosities are lower that was previously estimated. We found that
their highest luminosities do not exceed $\log L/L_{\sun}=5.1\pm0.2$
(see Miroshnichenko et al. \cite{m01}, \cite{m03}, \cite{m04}),
while previous studies, which used limited observations (e.g., with
no kinematical information), quoted higher values (e.g., 5.7 for MWC
300, Wolf \& Stahl \cite{ws85} and 6.0 for HDE 327083, Machado \&
Araujo \cite{ma03}).  At the same time, B[e] supergiants have not
been investigated thoroughly in the Milky Way. Thus observational
studies of known, but yet poorly-studied early-type stars with IR
excesses are important to reveal the population of dust-making
galactic hot stars.

In this paper we present and analyze optical and IR observations
obtained for two objects, \object{Hen\,3--298}\,=\,
IRAS\,09350$-$5314 and \object{Hen\,3--303}\,=\,IRAS\,09369$-$5406,
originally selected by Dong \& Hu (\cite{dong}) for their strong IR
excesses, but previously unstudied. The observations are described
in Section \ref{obs}, our new results are presented in Section
\ref{res}, our analysis of the observe properties is given in
Section \ref{discus}, and conclusions are summarized in Section
\ref{conclus}.

\section{Observations}\label{obs}

Our optical spectroscopic observations were obtained at the
2.1\,m telescope of the Complejo Astron\'omico El Leoncito (Argentina)
with the \'echelle spectrograph REOSC, mounted at the Cassegrain focus
and equipped with a 2000$\times$2000 pixel CCD chip. This setup allowed
us to achieve $R \sim$15\,000.  The log of our high-resolution
spectroscopic observations is presented in Table \ref{t1}. The data
reduction was performed in IRAF{\footnote{IRAF is distributed by the
National Optical Astronomy Observatories, which are operated by the
Association of Universities for Research in Astronomy, Inc., under
contract\ with the National Science Foundation.}}.

\begin{table}
\caption[]{Log of high-resolution spectroscopic observations of
Hen\,3--298 and Hen\,3--303. Listed are the object names, dates,
exposure starting times (in MJD=JD$-$2450000), exposure time in
seconds, spectral region in \AA, signal-to-noise ratios (at
5900\,\AA\ for the optical spectra)} \label{t1}
\begin{tabular}{cccrrr}
\hline\noalign{\smallskip}
Object & Date       & MJD      &$\!\!\!$Exp. &$\!\!\!$Sp. region&$\!\!\!$SNR\\
\noalign{\smallskip}\hline\noalign{\smallskip} Hen\,3--298&
$\!\!\!$2002/03/03 & $\!\!\!$2336.744 &$\!\!\!$ 1500 &
$\!\!\!$5600--8730 & $\!\!\!$40\\
Hen\,3--298& $\!\!\!$2002/04/15 & $\!\!\!$2379.582 &$\!\!\!$ 1800 &
$\!\!\!$5450--8415 & $\!\!\!$40\\
Hen\,3--298& $\!\!\!$2002/12/03 & $\!\!\!$2611.830 &$\!\!\!$ 180 &
$\!\!\!$23343-23450 & $\!\!\!$150\\
Hen\,3--298& $\!\!\!$2004/05/21 & $\!\!\!$3146.613 &$\!\!\!$ 720 &
$\!\!\!$23061-23179 & $\!\!\!$100\\
Hen\,3--298& $\!\!\!$2004/05/21 & $\!\!\!$3146.569 &$\!\!\!$ 720 &
$\!\!\!$23163-23279 & $\!\!\!$200\\
Hen\,3--298& $\!\!\!$2004/05/21 & $\!\!\!$3146.545 &$\!\!\!$ 720 &
$\!\!\!$23264-23378 & $\!\!\!$250\\
Hen\,3--298& $\!\!\!$2004/05/21 & $\!\!\!$3146.508 &$\!\!\!$ 720 &
$\!\!\!$23367-23479 & $\!\!\!$250\\
Hen\,3--303& $\!\!\!$2002/04/15 & $\!\!\!$2379.608 &$\!\!\!$ 1800 &
$\!\!\!$5450--8415 & $\!\!\!$20\\
Hen\,3--303&$\!\!\!$2002/05/22 & $\!\!\!$2417.473 &$\!\!\!$ 1500 &
$\!\!\!$5650--8750 & $\!\!\!$20\\
Hen\,3--303&$\!\!\!$2002/06/23 & $\!\!\!$2449.462 &$\!\!\!$ 1500 &
$\!\!\!$5650--8750 & $\!\!\!$20\\
Hen\,3--303&$\!\!\!$2002/12/03& $\!\!\!$2611.859 &$\!\!\!$ 720 &
$\!\!\!$23343-23450 & $\!\!\!$120\\
\noalign{\smallskip}\hline
\end{tabular}
\end{table}

Near-IR spectra were obtained at high-resolution ($R \sim$50000)
using the Gemini South telescope with the Phoenix spectrograph
(Hinkle et al.  \cite{h03}).  Spectra of both Hen\,3--298 and
Hen\,3--303 were obtained on 3 December 2002 in the 2.330 and 2.347
$\mu$m (4260 and 4290 cm$^{-1}$) region of the K band.  Another
spectrum of Hen\,3--298 was obtained on 21 May 2004, with the same
spectrograph, but with broader wavelength coverage from 2.304 to
2.347 $\mu$m (4260--4340 cm$^{-1}$).  A detailed description of the
observing and reduction process for Phoenix data taken at the same
resolution and wavelength can be found in Smith et al. (\cite{s02}).
As in Smith et al. (\cite{s02}), the telluric spectrum has been
ratioed from the program star spectra by reference to a featureless
hot star of similar airmass.  The use of several hot reference stars
and the radial velocity (RV) of the observed stellar features
confirms that the observed emission lines in Hen\,3--298 and
Hen\,3--303 are not absorption lines in the reference spectra.
Details of the observations can be found in Table \ref{t1}.

The near-IR broadband photometric observations were obtained at the
0.75--meter telescope of the South-African Astronomical Observatory
equipped with a single--element InSb photometer (Carter \cite{c90}).
Additionally $JHK$ data for the objects were extracted from the
2MASS catalog (Cutri et al. \cite{cutri}). The photometric data are
presented in Table \ref{t2}.
%and the second release of the DENIS survey (DENIS Consortium \cite{d03}).

%The DENIS $J$-band brightness of Hen\,3--298 agrees for the 2MASS and
%SAAO data within the measurement uncertainties, however, the DENIS
%$K$-band brightness measured at SAAO is $\sim$90\% higher.  Comparison
%of the 2MASS and DENIS data for the object's neigbourhood shows
%significant differences between the two data sets.  Since the 2MASS
%data for Hen\,3--298 are very close to those of SAAO in the other near-IR
%bands, we consider the DENIS $K$-band result a calibration issue and
%not a proof of the object's variability.

\begin{figure*}[htp]
\begin{center}
\resizebox{12cm}{!}{\includegraphics{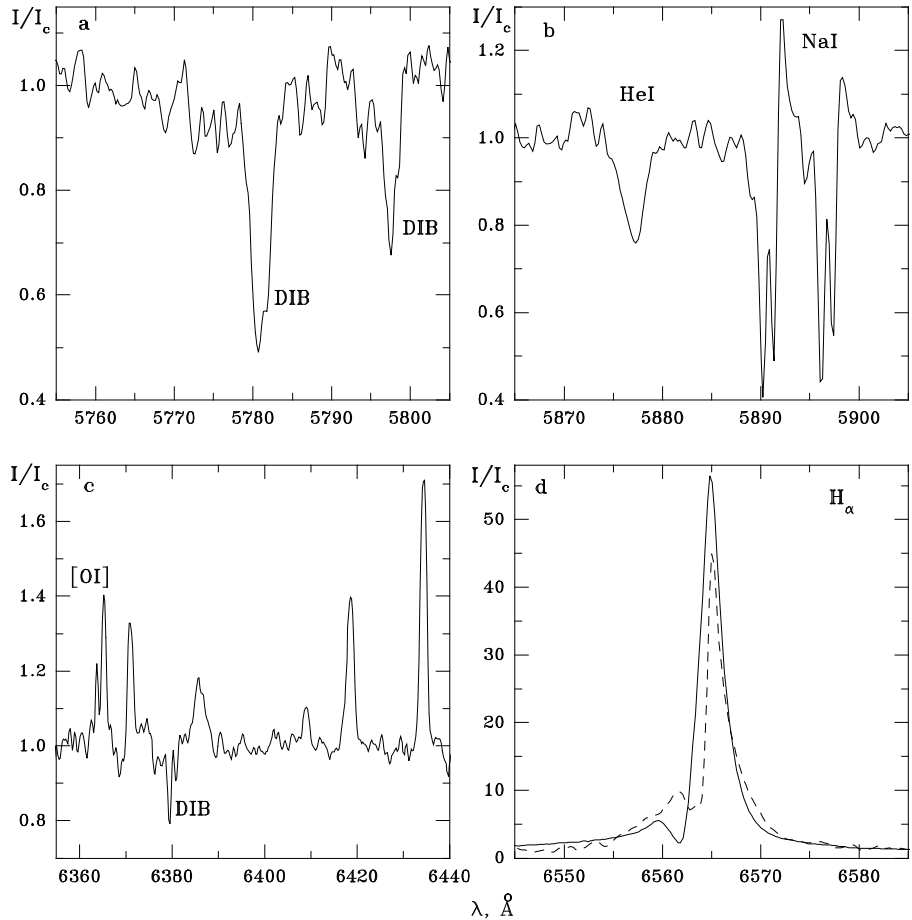}} \caption[]
{Portions of the optical spectra of Hen\,3--298 (solid lines). All
unmarked lines in panel c) are Fe {\sc II} lines. The H$\alpha$
profile of Hen\,3--303 is shown by the dashed line in panel d). The
intensities are in continuum units, and the wavelengths are in \AA.}
\label{f1}
\end{center}
\end{figure*}

The 2MASS survey detected 2 bright sources near the IRAS position of
Hen\,3--303 separated by $\sim 12\arcsec$. Both sources are present
in the GSC and USNO surveys with similar red magnitudes ($\sim$13.1
mag in the GSC red band). They both were included in the aperture of
the SAAO photometer (36$\arcsec$) and most likely were observed
together by IRAS. However, the MSX (Egan et al. \cite{e03}) detected
only one source with a high positional accuracy ($\sim 0\farcs4$).
It coincides with the fainter 2MASS object(09383462--5420260). Also,
the close IRAS and MSX fluxes suggest that the brighter 2MASS object
(09383455--5420144) does not contribute to the mid-IR (8--60 $\mu$m)
region. From its very red optical colour (the GSC blue--red
colour-index $\ge$4 mag) and smaller 2MASS colour-indices, one can
conclude that it is a cool star without a significant intrinsic IR
excess.

\begin{table*} \caption[]{Near-IR photometry of Hen\,3--298 and
Hen\,3--303. Listed are the object names, observing dates, magnitudes,
and sources of the data.} \label{t2} \begin{center}
\begin{tabular}{ccrcccl} \hline\noalign{\smallskip} Object   & JD
&$J$  & $H$  & $K$  & $L$  &Source\\
     &2400000+ &     &      &      &      &\\
\noalign{\smallskip}\hline\noalign{\smallskip}
Hen\,3--298&49120.29 &7.84$\pm$0.03 & 6.80$\pm$0.03 & 5.62$\pm$0.03 & 3.94$\pm$0.05 &SAAO\\
Hen\,3--298&51586.59 &7.85$\pm$0.03 & 6.84$\pm$0.06 & 5.70$\pm$0.02 & $-$           &2MASS\\
%Hen\,3--298&51827.98 &7.89$\pm$0.06 & $-$           & 4.87$\pm$0.09 & $-$           &DENIS\\
Hen\,3--298&52033.29 &7.95$\pm$0.03 & 6.87$\pm$0.03 & 5.66$\pm$0.03 & 3.98$\pm$0.05 &SAAO\\
\noalign{\smallskip}\hline\noalign{\smallskip}
Hen\,3--303$^{\rm a}$&49120.31 &8.46$\pm$0.03 & 7.05$\pm$0.03 &6.32$\pm$0.03 & 5.12$\pm$0.05 &SAAO\\
Hen\,3--303$^{\rm b}$&51553.79 &8.55$\pm$0.03 & 7.31$\pm$0.06 &6.77$\pm$0.02 & $-$           &2MASS\\
Hen\,3--303$^{\rm c}$&51553.79 &10.26$\pm$0.03& 8.89$\pm$0.06 &7.45$\pm$0.03 & $-$           &2MASS\\
Hen\,3--303$^{\rm a}$&52033.32 &8.48$\pm$0.03 & 7.06$\pm$0.03 &6.31$\pm$0.03 & 5.02$\pm$0.05 &SAAO\\
\noalign{\smallskip}\hline\noalign{\smallskip} \end{tabular}
\begin{list}{} \item $^{\rm a}$ -- combined photometry of the 2MASS
sources 09383455--5420144 and 09383462--5420260
\item $^{\rm b}$ -- source 09383455--5420144 at R.A. $9^{h}38^{m}34\fs56$,
 Dec.  $-54\degr20\arcmin14\farcs5$ (J2000)
\item $^{\rm c}$ -- source 09383462--5420260 at R.A. $9^{h}38^{m}34\fs63$,
Dec.  $-54\degr20\arcmin26\farcs0$ J2000) \end{list} \end{center}
\end{table*}

\begin{table*} \caption[]{Lines identified in the 2002/04/15 spectrum of Hen\,3--298.}
\label{t3}
\begin{center}
\begin{tabular}{lcrrrllcrrrll}
\hline\noalign{\smallskip} Line     & $\lambda_{\rm lab}$ & EW  & I/I$_{\rm c}$  & RV  & Comment &
 Line     & $\lambda_{\rm lab}$ & EW  & I/I$_{\rm c}$  & RV  & Comment\\
\noalign{\smallskip}\hline\noalign{\smallskip} DIB                 &
5780.41& 1.69& 0.49&  30.3&                 &Fe{\sc II}(74)       &
6456.38&  1.01& 1.68&    80.4& dp             \\ DIB
& 5849.80& 0.29& 0.70&  46.3&                 &Fe{\sc II}(40)
& 6516.05&  1.42& 1.91&    85.2&                \\ He{\sc I} (11)
& 5875.63& 0.61& 0.70&  99.7& noisy           &H {\sc I} (1)
& 6562.82&  0.00& 3.72&$-$137.6& blue peak      \\ Na{\sc I} (1)
& 5889.95& 0.73& 0.16&  15.4& blue comp       &H {\sc I} (1)
& 6562.82&140.00&39.10&   102.8& red peak       \\ Na{\sc I} (1)
& 5889.95& 0.38& 0.27&  75.0& red comp        &H {\sc I} (1)
& 6562.82&  0.00& 1.43&$-$43.5 & cd             \\ Na{\sc I} (1)
& 5889.95& 0.15& 1.35& 109.6& emis.           &DIB
& 6613.62&  0.52& 0.61&    15.8& blue comp.    \\ Na{\sc I} (1)
& 5895.92& 0.57& 0.20&  14.9& blue comp       &DIB
& 6613.62&  0.24& 0.78&    74.7& red comp.     \\ Na{\sc I} (1)
& 5895.92& 0.34& 0.24&  75.9& red comp        &$[$Fe{\sc
II}$]$(14F)& 7155.14&  1.22& 1.93&    85.0&               \\ Fe{\sc
II}(46)      & 5991.38& 0.80& 1.77&  74.1&                 &Fe{\sc
II}(73)       & 7224.51&  0.23& 1.19&    87.0&               \\
Fe{\sc II}(46)      & 6084.11& 0.32& 1.24&  75.9&
&$[$Ca{\sc II}$]$(1F) & 7291.46&  3.45& 3.59&    82.5& dp
\\ Fe{\sc II}(74)      & 6238.38& 0.96& 1.45& 101.7:&blend
&$[$Ca{\sc II}$]$(1F) & 7323.88&  2.31& 2.59&    80.8& dp
\\ Fe{\sc II}(74)      & 6247.56& 0.97& 1.48&  66.5:&blend
&Fe{\sc II}(73)       & 7449.34&  0.50& 1.24&    84.6& dp
\\ DIB                 & 6283.86& 2.20& 0.42&  37.5& noisy
&$[$Fe{\sc II}$]$(14F)& 7452.50&  0.51& 1.39&    87.0&
\\ $[$O{\sc I}$]$ (1F) & 6300.23& 0.91& 1.89&  85.4&
&Fe{\sc II}(73)       & 7462.38&  0.93& 1.42&    84.5& dp
\\ $[$O{\sc I}$]$ (1F) & 6363.88& 0.35& 1.31&  83.6&
&Fe{\sc II}(72)       & 7533.42&  0.39& 1.27&    72.5& dp
\\ Fe{\sc II} (40)     & 6369.45& 0.38& 1.31&  70.3& dp
&K{\sc I} (1)         & 7698.98&  0.33& 0.56&    15.8&
\\ Fe{\sc II} (40)     & 6369.45& 0.00& 1.27& 101.4&
&K{\sc I} (1)         & 7698.98&  0.13& 0.81&    86.4&
\\ DIB                 & 6379.20& 0.28& 0.74&  15.6&
&Fe{\sc II}(73)       & 7711.71&  1.55& 1.71&    83.5&
\\ Fe{\sc II}(74)      & 6407.30& 0.20& 1.15&  87.6& noisy
&H{\sc I} (11)        & 8323.43&  0.79& 1.30&    82.3&
\\ Fe{\sc II}(74)      & 6416.91& 0.70& 1.47&  82.8&
&H{\sc I} (11)        & 8333.79&  0.96& 1.31&    86.1&
\\ Fe{\sc II}(40)      & 6432.65& 1.12& 1.75&  70.5& dp
&H{\sc I} (11)        & 8345.55&  0.82& 1.38&    76.0& dp
\\ Fe{\sc II}(40)      & 6432.65& 0.00& 1.71&  98.9&
&H{\sc I} (11)        &
8359.01&  0.60& 1.35&    59.7& dp            \\
\noalign{\smallskip}\hline\noalign{\smallskip} \end{tabular}
\begin{list}{}
\item Line ID is listed in column 1, its laboratory
wavelength in column 2, equivalent width in \AA\ in column 3,
intensity in continuum units in column 4, heliocentric radial velocity
in km\,s$^{-1}$ in column 5, and a comment in column 6.
Columns 7--12 repeat the information.  Uncertain values are marked
with a colon.
\item dp stands for a double-peaked profile, cd for central depression
\end{list}
\end{center}
\end{table*}

\section{Results}\label{res}

\subsection{The spectrum}\label{res_sp}

Optical and IR high-resolution spectroscopic observations were
obtained for both objects for the first time. Optical spectra show
that both Hen\,3--298 and Hen\,3--303 have extremely strong
H$\alpha$ lines (Fig. \ref{f1}d) with double-peaked profiles and a
stronger red peak.  This is similar to most of the B[e]WD objects
and indicates the presence of a flattened gaseous envelope around
the objects.

The optical spectrum of Hen\,3--298 contains many emission lines of
Fe {\sc II} with mostly single-peaked profiles. Along with the
presence of the forbidden oxygen lines at 6300 and 6363 \AA\ (Fig.
\ref{f1}c) and the Ca {\sc II} IR triplet ($\lambda\lambda$ 8498,
8542, and 8662 \AA), this indicates that Hen\,3--298 is a typical
B[e] star. The detected He {\sc I} lines (at 5876 and 7065 \AA) are
in pure absorption, indicating a spectral type of no earlier than B3
(Fig.  \ref{f1}b). The width of the He {\sc I} 5876 \AA\ line (FWHM
$\sim$ 140 km\,s$^{-1}$) suggests that the star is not a rapid
rotator. The H$\alpha$ line varies in strength (the equivalent width
is 220 \AA\ in March and 140 \AA\ in April 2002), but has the same
position in our both spectra (central depression at $-44\pm$1
km\,s$^{-1}$) with a peak separation of 244$\pm$4 km\,s$^{-1}$. The
strong Na {\sc I} interstellar (IS) absorption lines and strong
diffuse IS bands (DIB, Fig. \ref{f1}ab) imply a high level of IS
reddening, supported by the available photometric data.

\begin{figure}[htp] \begin{center}
\resizebox{8cm}{!}{\includegraphics{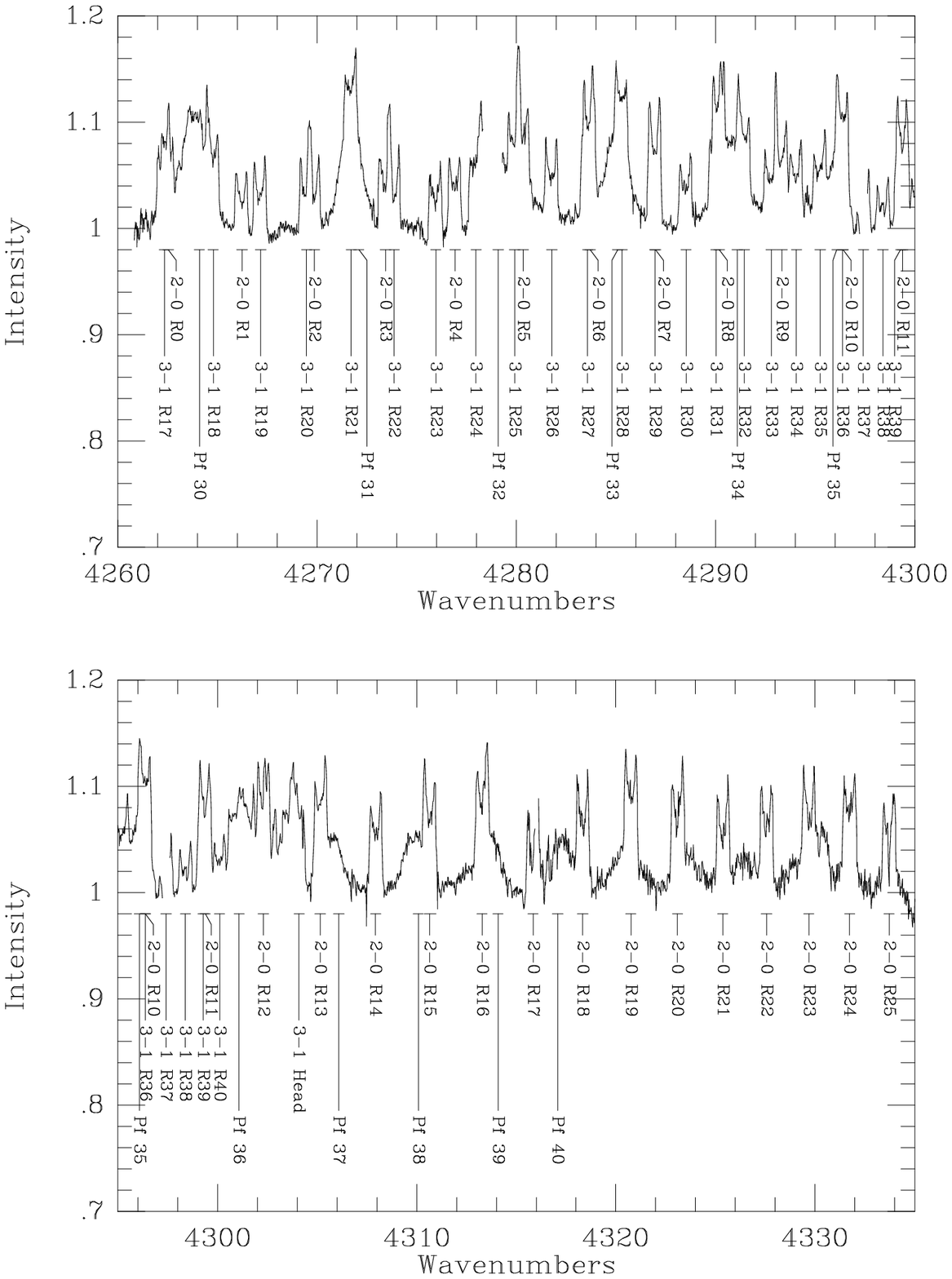}} \caption[] {The
IR spectrum of Hen\,3--298 observed at high resolution on May 21,
2004. The marked features are emission lines of the CO molecule
originating from the 2-0 and 3-1 bands. Positions of the hydrogen
lines of the Pfund series are also shown. The intensities are
normalized to the continuum and the frequency scale is in cm$^{-1}$.
In wavelength units the spectral region shown covers 2.3062 $\mu$m
(4335 cm$^{-1}$) to 2.3468 $\mu$m (4260 cm$^{-1}$).} \label{f2}
\end{center} \end{figure}

\begin{figure}[htp] \begin{center}
\resizebox{8cm}{!}{\includegraphics{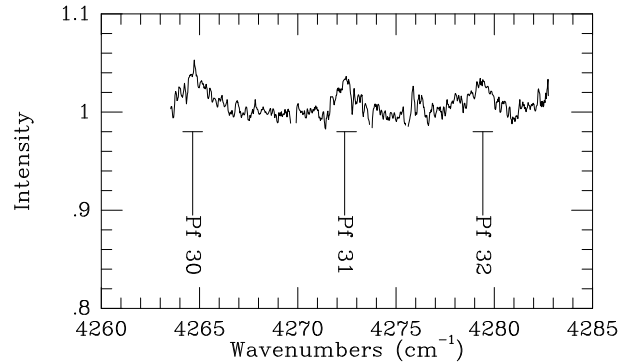}} \caption[] {The
IR spectrum of Hen\,3--303 observed at high resolution on December
3, 2002. The axes are as in Figure \ref{f2}. The hydrogen lines of
the Pfund series are labeled. In wavelength units the spectral
region shown covers 2.3343 $\mu$m (4283 cm$^{-1}$) to 2.3449 $\mu$m
(4263 cm$^{-1}$).} \label{f3}
\end{center} \end{figure}

\begin{figure}[htp] \begin{center}
\resizebox{8cm}{!}{\includegraphics{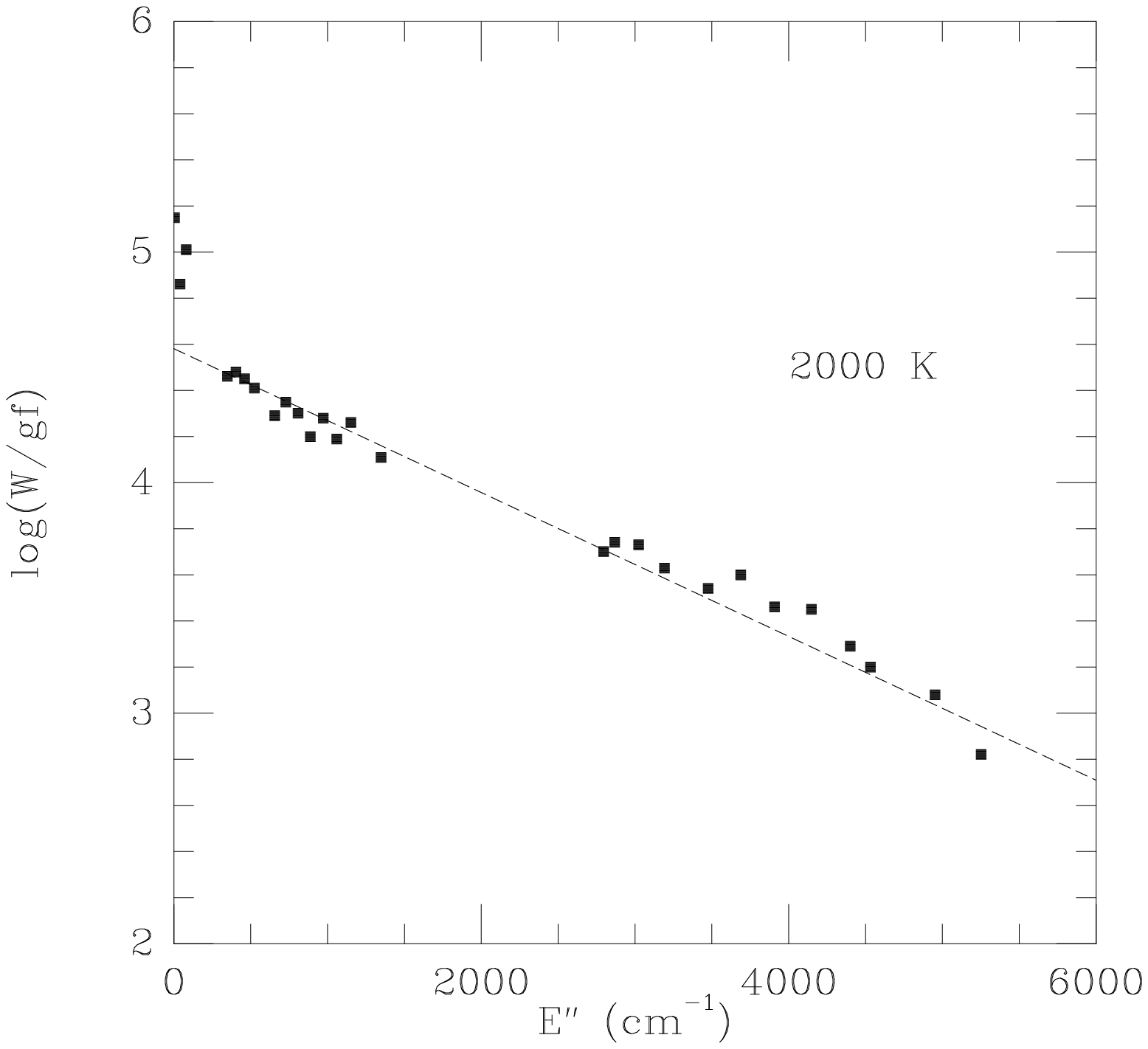}} \caption[]
{Excitation temperature of the CO lines in the spectrum of
Hen\,3--298. The line equivalent widths normalized to the product of
their statistical weight and oscillator strength are plotted versus
the energy level in cm$^{-1}$.} \label{f4} \end{center} \end{figure}

The IR spectrum (Fig. \ref{f2}) shows the presence of double-peaked
CO emission lines, supporting the disc-like density distribution of
the CS gas around Hen\,3--298.  Even a robust molecule line CO can
not survive in the photosphere of a B star, conclusively
demonstrating the CS origin of the CO emission line spectrum.  CO
lines can be identified from a number of vibration-rotational
levels.  Most of the lines from 2-0 R0 through 2-0 R 25 and 3-1 R17
through 3-1 R 40 can be identified.  Those that are not identified
are clearly present as blends.  Our spectra do not cover the region
from 2-0 R 26 to the 2-0 head at about R50 but the behaviour of the
3-1 lines and the strength of the observed 2-0 lines show that these
lines would be easily observable.

The observed CO lines cover a range in excitation energy, hence the
line strength normalized by the oscillator strength and statistical
weight can be plotted as a function of excitation energy to give the
rotational excitation temperature.  The line equivalent widths
suggest an excitation temperature of $\sim$2000 K (Fig. \ref{f4}).
This exceeds the condensation temperature of most dust, suggesting
that the CO spectrum originates in a dust free zone, near the dust
sublimation region of the disc. The lines peak separation can be
used to estimate the material rotational velocity (17.8$\pm$0.4
km\,s$^{-1}$).

Unfortunately, the optical spectra of Hen\,3--303, a fainter star,
have a lower SNR. No emission lines, except the H$\alpha$, Ca {\sc
II} IR triplet, and traces of Fe {\sc II} lines near 6400 \AA\ were
detected. The H$\alpha$ line shows strength variations from 180 \AA\
(April 2002) to 100 \AA\ (June 2002) with the same position (central
depression at 14$\pm$3 km\,s$^{-1}$ and a peak separation of
155$\pm$5 km\,s$^{-1}$). The Na {\sc I} IS absorptions are almost as
strong as those in Hen\,3--298 and also have two components (see
Fig. \ref{f1}b). No traces of CO bands were detected in the IR
spectrum of Hen\,3--303. The short interval of IR spectrum observed
(Figure \ref{f3}) does have the Pfund hydrogen lines in emission.
These lines are single peaked. From the three Pfund lines the
velocity is +55$\pm$2 km\,s$^{-1}$. The Paschen emission lines in
the spectrum of Hen\,3--303 show an RV of +80 km\,s$^{-1}$. The
difference in RV between the optical and IR lines may be a sign of
binarity. In any case, the combination of the SED of Hen\,3--303
with the strong H$\alpha$ emission suggest that it is also a B[e]
star, a member of the B[e]WD group.

We derived RVs of different spectral lines in the spectrum of
Hen\,3--298. The mean heliocentric RV of the Fe {\sc II} lines (from
both optical spectra, see Table \ref{t1}) is +83$\pm$3 km\,s$^{-1}$.
The He {\sc I} and Paschen emission lines show the same velocity.
The CO lines in the 2004 IR spectrum give the mean RV of +79$\pm$0.4
km\,s$^{-1}$, in agreement with the optical data.  Since the CO
lines sample both sides of a rotating disc around the star, these
lines should give an excellent measure of the stellar RV.

\subsection{Spectral energy distribution}\label{sed}

Henize (\cite{h76}) gives visual magnitudes of 10.4 and 12.5 for
Hen\,3--298 and Hen\,3--303, respectively.  Beyond this, optical
photometric data on the program stars are available only from
all-sky surveys (the Guide Star Catalog, Morrison et al. \cite{gsc};
the USNO--B1.0 catalog, Monet et al. \cite{usno}). The $R$--band
magnitudes from these surveys are 11.0 mag for Hen\,3--298 and 13.1
mag for Hen\,3--303. These optical brightnesses have uncertainties
of the order of 0.4 mag., but show a consistent magnitude
difference.

\begin{figure*}[t] \begin{center}
\resizebox{12cm}{!}{\includegraphics{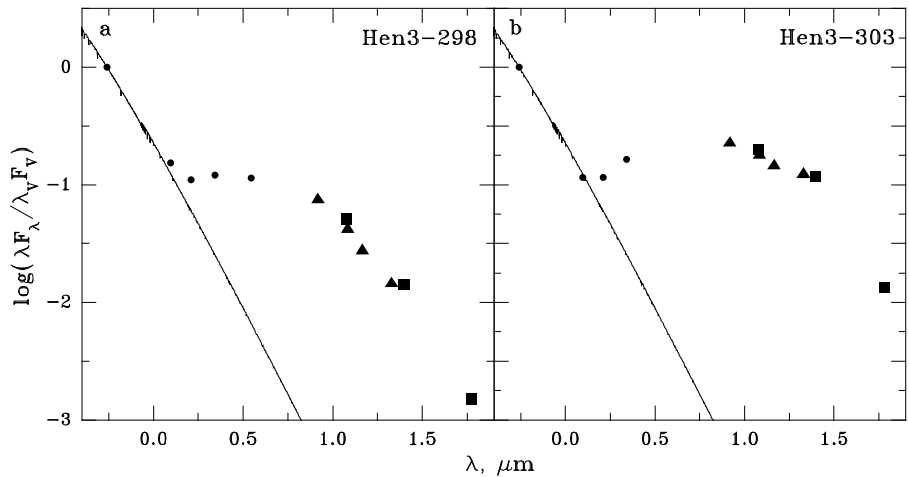}} \caption[] {The
observed spectral energy distributions of Hen\,3--298 (panel a) and
Hen\,3--303 (panel b) corrected for the IS extinction. The
ground-based data are shown by filled circles, the MSX data (Egan et
al. \cite{e03}) by filled triangles, the IRAS data by filled
squares. The solid line represents the Kurucz (\cite{k94})
theoretical model atmosphere for $T_{\rm eff}$=20000 K, $\log
g$=3.0). Refer to Sect. \ref{discus} for the IS reddening.}
\label{f5}
\end{center} \end{figure*}

We use the above visual magnitudes to normalize the SEDs shown in
Fig. \ref{f5}. The SEDs were corrected for the average galactic
extinction law (Savage \& Mathis \cite{sm79}) with a E($B-V$)=1.7
mag for both objects. The reddening correction was applied so that
the corrected $J$-band flux coincided with the theoretical model
atmosphere.  The accuracy of such a method is $\sim$0.1--0.2 mag in
E($B-V$) due to an uncertainty in the objects' spectral type and a
possible contribution from the CS gas.  However, experience with
other B[e]WD shows that this gives consistent results compared to
other methods (DIB strength, near-IR oxygen line ratio, e.g.,
Miroshnichenko et al. \cite{m02b}). Also, the IR excesses of both
objects turned out to be similar to those of other B[e]WD (see
Miroshnichenko et al. \cite{m01}).

There is a discrepancy of the derived E($B-V$) and that estimated from
the strength of the DIBs (at 5780 and 5797 \AA) in the spectrum of
Hen\,3--298 (E($B-V$) $\sim$3 mag) using a relationship from Herbig
(\cite{h93}). However, the relationship may not be linear for such
strong DIBs.  A different slope to this relation is also possible along
the line of sight to the program objects.  Thus, we consider the DIB
method less reliable in this case.

The IR photometry clearly shows that both objects have strong IR
excesses. The ground-based $JHKL$ photometry is consistent with both
the MSX and IRAS data (except in the case of the 2 sources observed
together at SAAO). The IRAS LRS spectrum (Olnon et al. \cite{o86})
of Hen\,3--298 is featureless. This is similar to those of most
other B[e]WD and may indicate the dusty envelope is disc-like and
have a large optical depth. The IRAS LRS spectrum of Hen\,3--303,
which is fainter in the 10--$\mu$m region than Hen\,3--298, is very
noisy.

\section{Discussion}\label{discus}

Our results show that both objects seem to be highly reddened B-type
stars with significant IR excesses due to CS dust. The high reddening
implies a large distance toward them. The IS extinction law in the
objects' direction (Neckel \& Klare \cite{nk80}) suggests an extinction
level of A$_{V} \sim$1 mag at distances $D \le$2 kpc followed by a
linear increase further away from the Sun, so that the A$_V$ exceeds 3
mag at $D \sim$4 kpc.

From the constructed SEDs and the fact that B[e]WD show almost no
excess radiation due to the CS envelope in the $J$--band, we estimated
the overall (CS and IS) extinction for Hen\,3--298 as A$_{V} \sim 5.4$
mag. Hen\,3--303 has a similar extinction. Part of the extinction may
arise in CS medium.

The RV data suggest a large distance toward the objects, if we
interpret them as due to the galactic rotation. According to Dubath,
Mayor, \& Burki (\cite{dmb88}), the heliocentric RV in this direction
is only 25 km\,s$^{-1}$ at 3 kpc. The rotation curve of these authors
is not applicable to farther distances from the Sun.

The two strong IS components of the Na {\sc I} doublet in the
spectrum of Hen\,3--298 and the two-component structure of some IS
features (e.g., the DIB at 6613 \AA\ and the K {\sc I} line at 7699
\AA) suggest the object may be located in the Norma spiral arm at
$D$=3--4.5 kpc. In this case the absolute visual magnitude of
Hen\,3--298 is M$_{V}= -6.8\pm$0.5. In combination with a mid-B
spectral type, it gives a M$_{\rm bol} \sim -$8 mag and a luminosity
of $\log L/L_{\sun} \sim$5.1. Thus, Hen\,3--298 seems to belong to
the highest luminosity subgroup of the B[e]WD, which contains such
objects as MWC\,300, HDE\,327083, CI Cam, and \object{Hen\,3--1398}.

The luminosity of Hen\,3--303 is less certain, as we were not able
to estimate its spectral type from our spectra. However, the
observed reddening and RV suggest that Hen\,3--303 is located in the
same spiral arm as Hen\,3--298. Taking into account the visual
magnitude of $\sim$2 mag fainter than that of Hen\,3--298, we can
estimate a luminosity of Hen\,3--303 to be $\log L/L_{\sun}
\sim$4.3. This estimate puts Hen\,3--303 into the
intermediate-luminosity subgroup of the B[e]WD, which includes
\object{MWC\,342}, \object{MWC\,623}, \object{MWC\,657},
\object{AS\,78}, \object{AS\,160}, and \object{HD\,85567}.

The H$\alpha$ line in both Hen\,3--298 and Hen\,3--303 has a central
depression, which is blueshifted from the systemic RV by $\sim$120
and $\sim$ 100 km\,s$^{-1}$, respectively. On the other hand, the CO
lines in the spectrum of Hen\,3--298 are narrow and symmetric with
respect to the systemic RV. This suggests that the CS gas
distribution is dominated by an outflow in the inner regions of the
gaseous envelope and by Keplerian rotation in the outer regions.
Thus, the stellar wind slows down significantly not far from the hot
star, where the material can accumulate, become dense enough to get
shielded from the stellar UV photons, and thus form dust.

The objects are very unlikely to be young stars because the
pre-main-sequence evolutionary time for B-type stars is too short
($\sim 10^{4}$--10$^{5}$ years, Palla \& Stahler \cite{ps93}) to get
rid of the distant (and thus cold) protostellar dust. Observations
of the most massive Herbig Be stars show that their IRAS fluxes at
60 and 100 $\mu$m are always larger than those at 12 and 25 $\mu$m.
Hen\,3--298 and Hen\,3--303 do not seem to be post-AGB objects
either due to reasons discussed by Miroshnichenko et al.
(\cite{m00}). Instead, they might recently left the main sequence
and begun to form dust.

The high luminosity of Hen\,3--298 and its strong emission-line
spectrum may suggest that the object is a single B[e] supergiant, in
whose slow and dense wind the dust may form (e.g., Kraus \& Lamers
\cite{kl03}). On the other hand, there is strong evidence that most
galactic B[e] supergiants are binary systems. The examples include
recognized binaries, such as \object{RY Sct} (e.g., Gehrz et al.
\cite{g01}) and HDE 327083 (Miroshnichenko et al. \cite{m03}), and a
suspected one (MWC 300, Miroshnichenko et al. \cite{m04}).

It is harder to explain the existence of a strong wind in a
lower-luminosity Hen\,3--303, so that binarity might be more likely
for it. Our previous results on B[e]WD (e.g., Miroshnichenko
\cite{m00}, \cite{m02b}) show that the secondaries are 2--3 mag
fainter than the primaries in the optical region and strongly call
for additional high signal-to-noise spectroscopy of Hen\,3--303. The
RV difference in the optical and IR spectra, taken not
contemporaneously, might be due to orbital motion. A lower systemic
velocity does not contradict the location of the object in the Norma
spiral arm. Also, the very strong H$\alpha$ emission is consistent
with a B spectral type of the underlying star. Although our data are
not sufficient to constrain its temperature, stars of later spectral
types are unlikely to display such an emission-line spectrum.

\section{Conclusions}\label{conclus}

Summarizing the results reported above, we can draw the following
conclusions about the properties, nature, and evolutionary state of the
emission-line objects Hen\,3--298 and Hen\,3--303.

\begin{enumerate}
\item Our high-resolution optical spectra show that
both objects have strong line emission, similar to those of B[e]WD. The
RVs of most of the emission and absorption lines suggest a large
distance toward them. In combination with the photometric data, this
suggests that the objects are located in the Norma spiral arm at a
distance of 3--4.5 kpc.
\item The double-peaked profiles of
optically-thick spectral lines (e.g., H$\alpha$) suggest a disc-like
distribution of the CS gas near both objects.  The double-peaked CO
line profiles seen in the spectrum of Hen\,3--298 also suggest such
a distribution for the molecular part of the envelope.  The
featureless IRAS LRS of Hen\,3--298 in the 10--$\mu$m region is
consistent with a large optical depth of the dusty envelope, which
most likely has a disc-like shape. The discs in both objects are
viewed close to edge-on.  The inclination of Hen\,3--303 may be
slightly steeper than that of Hen\,3--298, so that the dust obscures
the inner dissociation ring where CO originates.
\item Our estimate of the luminosity of Hen\,3--298 ($\log
L/L_{\sun} \sim$5.1) suggests that it is among the highest
luminosity galactic B[e]WD. Hen\,3--303, with its smaller optical
brightness, probably has a lower luminosity ($\log L/L_{\sun}
\sim$4.3).
\end{enumerate}

As it was shown by Miroshnichenko et al. (\cite{m00}), (\cite{m01}),
(\cite{m02b}), (\cite{m03}), and (\cite{m04}), almost all of the
quoted above objects from the highest- and intermediate-luminosity
subgroups of B[e]WD are either confirmed or suspected binaries.
Thus, it is worthwhile to look for signs of binarity in both
Hen\,3--298 and Hen\,3--303.

The following observations need to be performed in order to place
stronger constraints on the objects' parameters. Optical photometry is
needed to constrain the reddening and luminosity. High-resolution
spectroscopy with higher SNR is needed to constrain spectral types and
search for RV variations.  Moderate resolution IR spectroscopy in the
10--$\mu$m region could reveal the chemical composition of the CS dust.

\section{Acknowledgments}
We thank T. Lloyd Evans for obtaining some of the near-IR
photometric data at the SAAO and making them available to us. A.~M.
and K.~S.~B. acknowledge support from NASA grant NAG5--8054. We
thank Dr. Robert Blum for assistance with the May 2004 Phoenix
observations. This paper is based in part on observations obtained
at the Gemini Observatory, which is operated by the Association of
Universities for Research in Astronomy, Inc., under a cooperative
agreement with the NSF on behalf of the Gemini partnership: the
National Science Foundation (United States), the Particle Physics
and Astronomy Research Council (United Kingdom), the National
Research Council (Canada), CONICYT (Chile), the Australian Research
Council (Australia), CNPq (Brazil), and CONICRT (Argentina).  The
observations were obtained with the Phoenix infrared spectrograph,
which was developed and is operated by the National Optical
Astronomy Observatory.  The spectra were obtained through programs
GS-2002B-Q-34 and GS-2004A-DD-1. This research has made use of the
SIMBAD database operated at CDS, Strasbourg, France, as well as of
data from the Two Micron All Sky Survey, which is a joint project of
the University of Massachusetts and the Infrared Processing and
Analysis Center/California Institute of Technology, funded by the
National Aeronautics and Space Administration and the National
Science Foundation.

\listofobjects


\begin{thebibliography}{}

\bibitem[1976]{as76}
Allen, D.A., \& Swings, J.-P. 1976, A\&A, 47, 293

\bibitem[1998]{b98}
Bjorkman, J.E. 1998, in B[e] stars, (eds.) A.-M.~Hubert and
C.~Jaschek, Kluwer Acad. Publ., p. 189

\bibitem[1990]{c90}
Carter, B.S. 1990, MNRAS, 242, 1

\bibitem[2003]{cutri}
Cutri, R.M., Skrutskie, M.F., van Dyk, S., et al. 2003,
CDS/ADC Collection of Electronic Catalogues, 2246

\bibitem[1991]{dong}
Dong, Y.S., \& Hu, J.Y. 1991, Chin. A\&A, 15, 275

\bibitem[1988]{dmb88}
Dubath, P., Mayor, M., \& Burki, G. 1988, A\&A, 205, 77

\bibitem[2003]{e03}
Egan, M.P., Price, S.D., Kraemer, K.E., et al.
2003, The Midcourse Space Experiment Point Source Catalog Version 2.3
(October 2003), AFRL--VS--TR--2003--1589

\bibitem[2001]{g01}
Gehrz, R.D., Smith, N., Jones, B., Puetter, R., \& Yahil, A. 2001,
ApJ, 559, 395

\bibitem[1976]{h76}
Henize, K.G. 1976, ApJS, 30, 491

\bibitem[1993]{h93}
Herbig, G.H. 1993, ApJ, 407, 142

\bibitem[2003]{h03}
Hinkle, K.H., Blum, R.D., Joyce, R.R., et al. 2003, SPIE, 4834, 353

\bibitem[2003]{kl03}
Kraus, M., \& Lamers, H.J.G.L.M. 2003, A\&A, 405, 165

\bibitem[1994]{k94}
Kurucz, R.L. 1994, Smithsonian Astrophys. Obs., CD-ROM No. 19

\bibitem[1998]{l98}
Lamers, H.J.G.L.M., Zickgraf, F.-J., de Winter, D., Houziaux,
L., \& Zorec, J. 1998, A\&A, 340, 117

\bibitem[2003]{ma03}
Machado, M.A.D., \& Ara\'ujo, F.X. 2003, A\&A, 409, 665

\bibitem[2000]{m00}
Miroshnichenko, A.S., Chentsov, E.L., Klochkova, V.G., et al. 2000,
A\&AS, 147, 5

\bibitem[2001]{m01}
Miroshnichenko, A.S., Levato, H., Bjorkman, K.S., \& Grosso, M.
2001, A\&A, 371, 600

\bibitem[2002a]{m02a}
Miroshnichenko, A.S., Bjorkman, K.S., Chentsov, E.L., \& Klochkova,
V.G., 2002a, in Exotic Stars as Challenges to Evolution, (eds.) C.A.
Tout and W. Van Hamme, ASP Conf.  Ser., 279, 303

\bibitem[2002b]{m02b}
Miroshnichenko, A.S., Bjorkman, K.S., Chentsov, E.L., et al. 2002b,
A\&A, 383, 171

\bibitem[2003]{m03}
Miroshnichenko, A.S., Levato, H., Bjorkman, K.S., \& Grosso, M.
2003, A\&A, 406, 673

\bibitem[2004]{m04}
Miroshnichenko, A.S., Levato, H., Bjorkman, K.S., et al. 2004, A\&A,
417, 731

\bibitem[2003]{usno}
Monet, D.G., Levine, S.E., Canzian, B., et al. 2003, AJ, 125, 984

\bibitem[2001]{gsc}
Morrison, J.E., Roeser, S., McLean, B., Bucciarelli, B., \&
Lasker, B. 2001, AJ, 121, 1752

\bibitem[1980]{nk80}
Neckel, Th., \& Klare, G. 1980, A\&AS, 42, 251

\bibitem[1986]{o86}
Olnon, F.M., Raymond, E., and the IRAS Science Team 1986, A\&AS, 65, 607

\bibitem[1993]{ps93}
Palla, F., \& Stahler, S.W. 1993, ApJ, 418, 414

\bibitem[1979]{sm79}
Savage, B.D., \& Mathis, J.S. 1979, ARA\&A, 17, 73

\bibitem[2002]{s02}
Smith, V.V., Hinkle, K.H., Cunha, K., et al. 2002, AJ, 124, 3241

\bibitem[1994]{twp94}
Th\'e, P.S., de Winter, D., \& P\'erez, M.R.  1994, A\&AS, 104, 315

\bibitem[1970]{wack}
Wackerling, L.R. 1970, Mem. RAS, 73, 153

\bibitem[1985]{ws85}
Wolf, B., \& Stahl, O., 1985, A\&A, 148, 412

\bibitem[1986]{z86}
Zickgraf, F.-J., Wolf, B., Leitherer, C., Appenzeller, I., \& Stahl,
O.  1986, A\&A, 163, 119

\end{thebibliography}
\end{document}